\documentclass[twocolumn]{aastex7}

\shorttitle{reproducing the SO/PHI observation}
\shortauthors{Wang et al.}

\graphicspath{{./}{figures/}}

\usepackage{CJK}
\usepackage{multirow}
\usepackage{booktabs}
\usepackage{amsmath}
\usepackage{threeparttable}
\usepackage{verbatim}

\usepackage{hyperref}

\begin{document}
\begin{CJK*}{UTF8}{gbsn}

\title{Successfully Reproducing Solar Orbiter Polar Field Observations with a Surface Flux Transport Model}

\correspondingauthor{Jie Jiang}
\email{jiejiang@buaa.edu.cn}

\author[orcid=0000-0003-0256-8102,sname=Ruihui,gname=Wang]{Ruihui Wang (王瑞慧)}
\affiliation{School of Space and Earth Sciences, Beihang University, Beijing, People’s Republic of China}
\affiliation{Key Laboratory of Space Environment Monitoring and Information Processing of MIIT, Beijing, People’s Republic of China}
\email[show]{wangruihui@buaa.edu.cn} 

\author[orcid=0000-0001-5002-0577,sname=Jie,gname=Jiang]{Jie Jiang (姜杰)}
\affiliation{School of Space and Earth Sciences, Beihang University, Beijing, People’s Republic of China}
\affiliation{Key Laboratory of Space Environment Monitoring and Information Processing of MIIT, Beijing, People’s Republic of China}
\email[show]{jiejiang@buaa.edu.cn}  

\author[orcid=0000-0002-6977-1239,sname=Yukun,gname=Luo]{Yukun Luo (罗昱琨)}
\affiliation{School of Physical Science and Technology, Southwest Jiaotong University, Chengdu 611756, People’s Republic of China;}
\affiliation{Astrophysical Center, Southwest Jiaotong University, Chengdu 611756, People’s Republic of China}
\email{luoyukun@buaa.edu.cn}  

\author[orcid=0000-0003-2755-5295,sname=D.,gname=Calchetti]{D. Calchetti}
\affiliation{Max-Planck-Institut f\"ur Sonnensystemforschung, Justus-von-Liebig-Weg 3, 37077 G\"ottingen, Germany}
\email{calchetti@mps.mpg.de} 

\author[orcid=0000-0002-3418-8449,sname=S.K., gname=Solanki]{S. K. Solanki}
\affiliation{Max-Planck-Institut f\"ur Sonnensystemforschung, Justus-von-Liebig-Weg 3, 37077 G\"ottingen, Germany}
\affiliation{School of Space Research, Kyung Hee University, Yongin, Gyeonggi 17104, Republic of Korea}
\email{solanki@mps.mpg.de}

\begin{abstract}
The solar polar field plays a fundamental role in predicting solar-cycle strength and determines the large-scale structure of the heliospheric magnetic field. The High Resolution Telescope of the Polarimetric and Helioseismic Imager aboard Solar Orbiter (SO/PHI-HRT) provides unprecedented observations of the solar polar regions, as Solar Orbiter currently reaches heliographic latitudes of up to $\pm17^\circ$, about $10^\circ$ higher than those accessible from near-ecliptic observations. Using these observations to further test surface flux transport models, we present a successful simulation of the SO/PHI-HRT observations, reproducing the latitudinal profiles of the longitudinally averaged radial magnetic field in both polar regions. This result is achieved using the meridional-flow profile proposed by \citet{WangRH2026}, which also reproduces the evolution of the axial dipole strength, magnetic butterfly diagram, and high-latitude magnetic fields observed by SDO/HMI during Solar Cycle 25. We further compare simulations using this profile with simulations employing two representative meridional-flow profiles characterized by different peak latitudes. The latter fail to reproduce the observed high-latitude and polar fields. The comparison demonstrates that the meridional-flow profile strongly regulates poleward flux transport and the resulting polar-field evolution. Profiles peaking at higher latitudes produce stronger flux surges, and stronger and more concentrated polar fields. These results provide new constraints on the large-scale solar meridional flow.

% The success of this profile suggests that the large-scale meridional flow on the solar surface may possess a steeper equatorial gradient and weaker mid-latitude flow speed than assumed in conventional SFT models.

\end{abstract}

\keywords{\uat{Solar physics}{1476}---\uat{Solar cycle}{1487}---\uat{Solar magnetic fields}{1503}---\uat{Solar surface}{1527}}

\section{Introduction} \label{sec:intro}

Solar polar fields are a fundamental component of the Sun's large-scale magnetic field. They dominate the large-scale structure of the corona, supply most of the interplanetary magnetic field, and regulate the fast solar wind \citep{wangYM1990, Owens2013, Petrie2015}. In addition, the polar-field strength around solar minimum is widely regarded as one of the most reliable precursors of the amplitude of the following solar cycle \citep{Schatten1978, Petrovay2020SCpd, Karak2026}. 

Observations of the solar polar field are, however, strongly limited by the near-ecliptic viewing angle. From Earth and near-earth spacecrafts, the poles can only be observed with maximum heliographic latitude of approximately $7.25^\circ$, corresponding to viewing angles exceeding $82^\circ$ \citep{Sun2011, Deng2025ChJSS}. The large viewing angle results in severe foreshortening, low spatial resolution, and increased measurement uncertainties \citep{Petrie2015}. Beginning in February 2025, the orbit of Solar Orbiter (\citealt{Muller2020}) enabled observations from heliographic latitudes up to approximately $17^\circ$, substantially improving the visibility of the polar regions compared with observations from the ecliptic plane. The first polar-field measurements obtained by the High Resolution Telescope of the Polarimetric and Helioseismic Imager onboard Solar Orbiter (SO/PHI-HRT; \citealt{Gandorfer2018, Solanki2020}) were reported by \citet{Calchetti2025}. These observations provide the most accurate data of the solar polar magnetic field currently available and offer a valuable opportunity to research the polar fields.

The evolution of the solar surface magnetic field can be successfully described by the surface flux transport (SFT) model \citep{Jiang2014ssr, Yeates2023}. In the model, newly emerged active regions (ARs) provide the source of magnetic flux, which is subsequently transported by differential rotation, meridional flow, and supergranular diffusion. While most of the emerged flux is canceled through interactions with opposite-polarity flux, a small fraction is transported poleward and contributes to the evolution of the polar fields. This transport is often episodic, producing prominent poleward-moving flux flows known as surges or plumes. SFT models have successfully reproduced many observed properties of the solar surface magnetic field, including the axial dipole strength, magnetic butterfly diagrams, and magnetic power spectra \citep{Cameron2010,Bhowmik2018, 2020Yeates, Jha2024, Upton2024, Wang2025, Luo2025}.

Compared with the axial dipole strength and other global quantities, the polar field itself remains less extensively studied in SFT simulations. Although many simulations successfully reproduce the timing of polar-field reversals and the overall long-term evolution \citep{Jiang2011two,Iijima2019,Yeates2025}, properly reproducing the observed polar-field strength and variability remains challenging \citep{WangYM1989ApJ, Virtanen2017, YangSH2024, YangD2024}. One major difficulty arises from the strong sensitivity of the polar field to the meridional-flow profile. \citet{DeVore1984} showed that the rate at which the meridional flow declines toward the poles strongly influences the latitudinal concentration of the polar field. \citet{DeVore1987} further demonstrated that broader flow profiles and profiles peaking at higher latitudes tend to produce stronger polar fields. Similarly, \citet{WangYM2017} found that meridional flows peaking at middle latitudes generate stronger surges and polar fields than profiles peaking closer to the equator.

Despite its importance, the meridional flow at the solar surface remains poorly constrained observationally \citep{Jiang2023}. To better reproduce the observed polar-field evolution, \citet{WangRH2026} proposed a new meridional-flow profile based on the functional form introduced by \citet{Lemerle2015}, with its parameters optimized using SFT simulations and high-latitude field observations from the Helioseismic and Magnetic Imager onboard the Solar Dynamics Observatory (SDO/HMI; \citealt{HMI}). The resulting profile successfully reproduces the evolution of both the axial dipole strength and the polar fields during 2020--2026 and successfully predicts the exceptionally weak northern polar field observed in 2026.

In this paper, we use the meridional-flow profile proposed by \citet{WangRH2026} to simulate the recent polar-field observations from SO/PHI-HRT. We investigate whether a single SFT simulation can simultaneously reproduce the out-of-ecliptic polar-field measurements from SO/PHI-HRT and the large-scale surface magnetic field evolution observed by SDO/HMI. We further compare simulations using this profile with those employing two representative meridional-flow profiles to investigate how the meridional-flow profile regulates poleward flux transport and the evolution of the solar polar magnetic field.

This paper is organized as follows. Section~\ref{sec:model} describes the SFT model and the three meridional-flow profiles used in this study. Section~\ref{sec:results} presents comparisons between the simulations and the HMI and SO/PHI-HRT observations. Finally, Section~\ref{sec:conclusion} summarizes the main results.

% Polar field simulation and the impact of meridional flow on polar field is less researched. 
% use the SO/PHI-HRT data test the SFT model, and new flow. more data with high quality will be given in future. simulation of polar field to test model performance, with or without other scheme. constrain meridional flow.

% evolution of surface field, surface flux 
% AR source to the polar field simulation; simply introduction

\section{Model Description}\label{sec:model}

The surface flux transport (SFT) model is governed by 
\begin{equation}
\label{eq:SFT}
\begin{split}
\frac{\partial B}{\partial t} &= -\omega(\theta) \frac{\partial B}{\partial \phi} - \frac{1}{R_\odot \sin \theta} \frac{\partial}{\partial \theta} [u(\theta) B \sin \theta] \\
&+ \frac{\eta}{R_\odot^2} \left[ \frac{1}{\sin \theta} \frac{\partial}{\partial \theta} \left( \sin \theta \frac{\partial B}{\partial \theta} \right) + \frac{1}{\sin^2 \theta} \frac{\partial^2 B}{\partial \phi^2} \right]\\
&+ S(\theta, \phi, t),
\end{split}
\end{equation}
where $B$ is the radial magnetic field, $\theta$ is the colatitude, $\phi$ is the longitude, $\eta$ is the supergranular diffusivity, and $u$ is the meridional flow. The differential rotation profile $\omega(\theta)$ is adopted from \citet{Snodgrass1983}. 

The meridional flow remains poorly constrained observationally. To improve the polar-field simulation, we proposed a new meridional-flow profile in \citet{WangRH2026}, based on the functional form introduced by \citet{Lemerle2015}. This profile, hereafter referred to as Flow 2, is given by
\begin{equation}
\label{eq:MF_new}
u(\theta) = u_0 \operatorname{erf}^8(2.25 \sin \theta) \operatorname{erf}(3.5 \cos \theta),
\end{equation}
where $u_0 = 13~\mathrm{m\ s^{-1}}$.
To better illustrate the influence of the meridional flow on the surface magnetic-field simulation, we also implement two representative flow profiles from \citet{van_Ballegooijen1998} and \citet{WangYM2017}, denoted as Flow 1 and Flow 3, respectively. Flow 1 is expressed as
\begin{equation}
\label{eq:MF_VB}
u(\lambda) = 
\begin{cases}
u_0 \sin \left( \pi \lambda / \lambda_0 \right) & \text{if } |\lambda| < \lambda_0 \\
0 & \text{otherwise},
\end{cases}
\end{equation}
where $\lambda$ is the latitude, $\lambda_0=75^\circ$  and $u_0 = 15~\mathrm{m\ s^{-1}}$.
Flow 3 is given by
\begin{equation}
\label{eq:MF_YM}
v(\lambda) = 1.08\ u_0 \tanh \left( \lambda / 6^\circ \right) \cos^2 \lambda,
\end{equation}
where $u_0 = 8~\mathrm{m\ s^{-1}}$.

% Compared to Flow 2, Flow 1 is slower between roughly $20 ^{\circ}$ and $70 ^{\circ}$ latitude. Its peak speed is both lower in magnitude and shifted equatorward, occurring at $29^{\circ}$. The equatorial latitudinal gradient ($D_u$) of the new profile is about $0.9 m\,s^{-1}\,{deg}^{-1}$ for $u_0 = 13$ m/s, which is consistent with most helioseismology measurements \citep{Zhao2014, Gizon2020, Jiang2023} and steeper than that of Flow 2. A comparison of their simulation results is presented in Section \ref{sus:postdiction}. 

\begin{figure}[htbp!]
\centering
\includegraphics[scale=0.33]{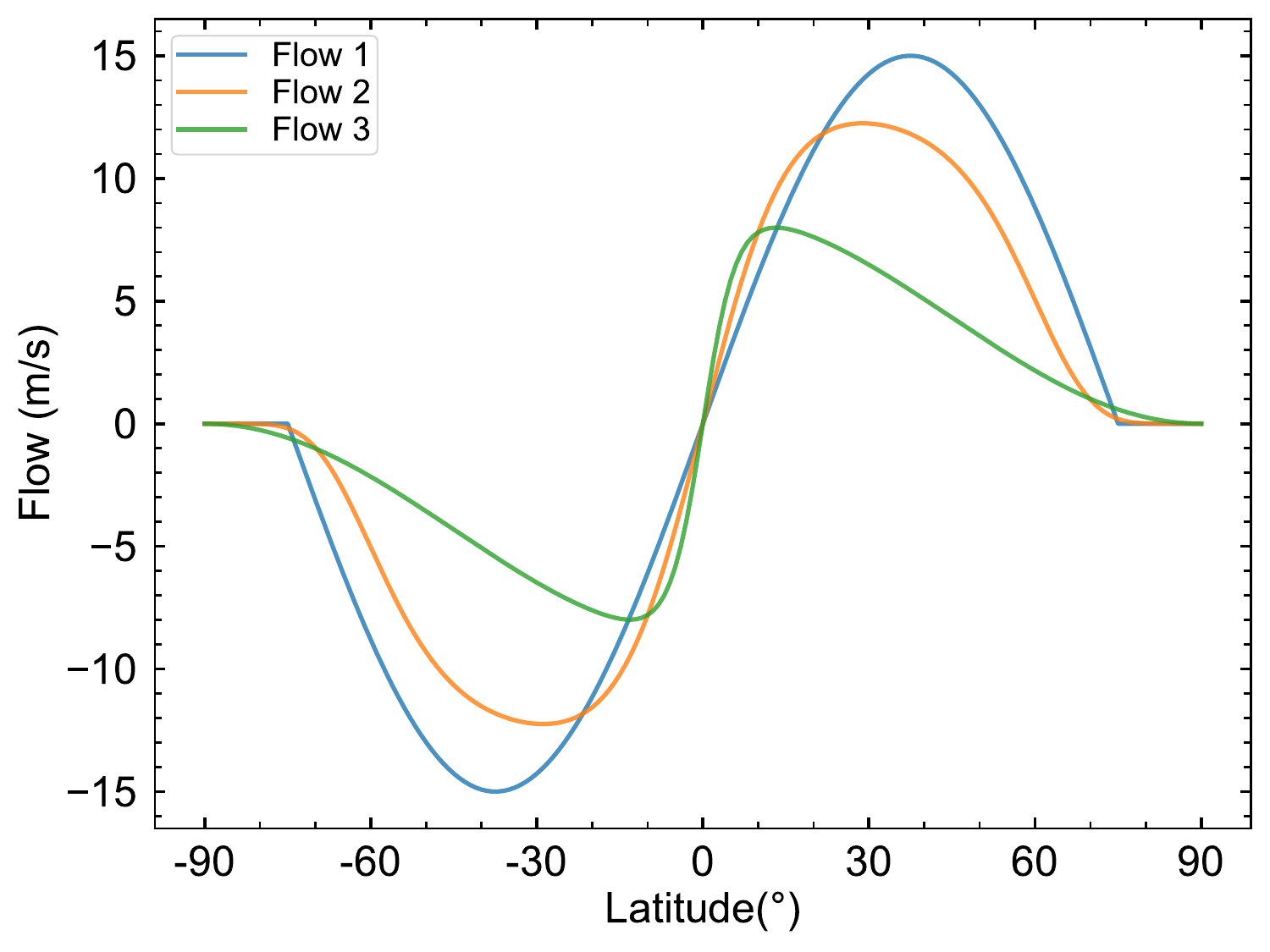}
\caption{Comparison of the three meridional flow profiles. Flow 1 follows the profile of \citet{van_Ballegooijen1998}. Flow 2 corresponds to the profile proposed by \citet{WangRH2026}. Flow 3 follows the profile of \citet{WangYM2017}.
\label{fig:MF}}
\end{figure}

Both the meridional flow speed of each flow ($u_0$) and supergranular diffusivity ($\eta$) are free parameters in the SFT simulation. Since the supergranular diffusivity and the equatorial latitudinal gradient of the meridional flow (controlled by $u_0$) largely govern the evolution of the axial dipole strength \citep{Petrovay2020AM}, to enable a fair comparison among the three flow profiles, we adjust the diffusivity and meridional flow speed for each case such that all simulations produce identical global axial dipole strengths. The meridional flow speed of each flow is given in equations~\eqref{eq:MF_new}, \eqref{eq:MF_VB}, and \eqref{eq:MF_YM}. For the supergranular diffusivity, we adopt $\eta = 300~\mathrm{km^2\ s^{-1}}$ for Flow 1 and $\eta = 450~\mathrm{km^2\ s^{-1}}$ for both Flow 2 and Flow 3. 

The three meridional-flow profiles with adjusted maximum flow speed are compared in Figure~\ref{fig:MF}. From Flow 1 to Flow 3, the latitude of the peak flow speed shifts progressively toward the equator. Meanwhile, the flow speed within approximately $20^\circ$--$70^\circ$ latitude decreases systematically, whereas the latitudinal gradient of the meridional flow at the equator ($D_u$) increases. 

The source term $S(\theta, \phi, t)$ is constructed from ARs in the Active Region database for Influence on Solar cycle Evolution (ARISE; \citealt{Wang2023, Wang2024}). Version~4.0 of the database and the associated codes are archived in Zenodo \citep{wang_2026_database}. The database does not include all small ARs, which may introduce some uncertainty into the simulations. However, the importance of small ARs for polar-field evolution remains under debate. \citet{Hofer2024} suggested that they may contribute significantly, whereas \citet{Whitbread2018} and \citet{Wang2024} found that reproducing the observed polar-field evolution does not require a complete inventory of small ARs. The capability of ARISE has been demonstrated through continuous SFT simulations from 2010 to 2024, which successfully reproduce the large-scale evolution of the surface magnetic field \citep{Wang2025}.

The SFT equation is solved using our newly developed code that employs a spectral method \citep{Wang2025, Luo2025}. The accuracy of the code has been validated through successful reproductions of the magnetic power spectra, axial dipole evolution, and magnetic butterfly diagram \citep{Wang2025, Luo2025,WangRH2026}. The magnetic field is expanded up to spherical harmonic degree $\ell_{\rm max}=60$, and both the input and output magnetograms are represented on a $180 \times 360$ grid.

\section{Results}\label{sec:results}

\subsection{Simulation of the ongoing cycle 25}\label{subs:sml_cycle25}

\begin{figure}[htbp]
\centering
\includegraphics[scale=0.33]{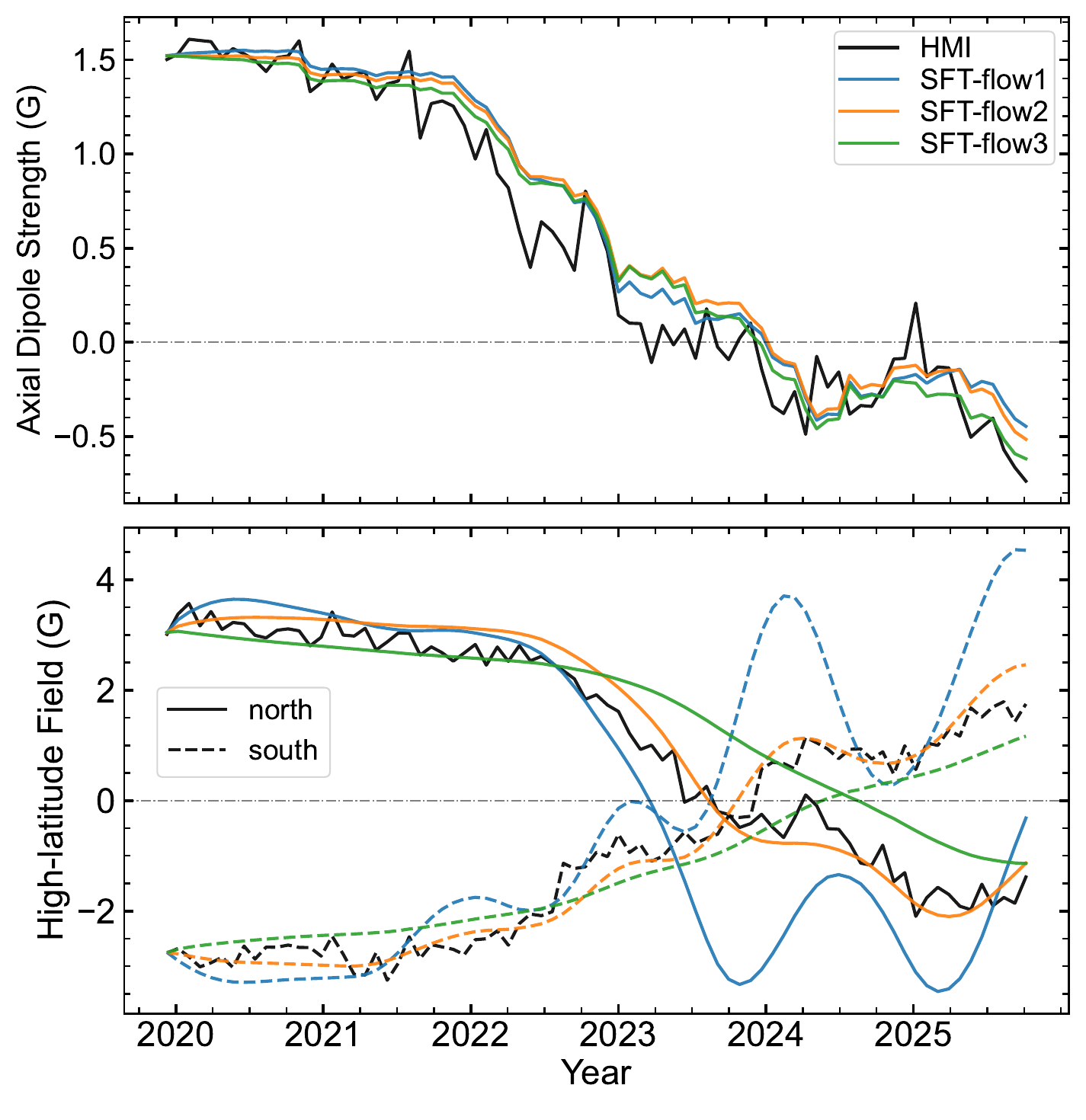}
\caption{Comparison of the observed and simulated axial dipole strength and high-latitude field during 2020--2025. The simulations employ the three meridional-flow profiles defined by Equations~\eqref{eq:MF_new}, \eqref{eq:MF_VB}, and \eqref{eq:MF_YM}. The high-latitude field is defined as the mean radial magnetic field over the latitude range $60^\circ$--$75^\circ$.
\label{fig:df_pf}}
\end{figure}

\begin{figure*}[htbp]
\centering
\includegraphics[scale=0.5]{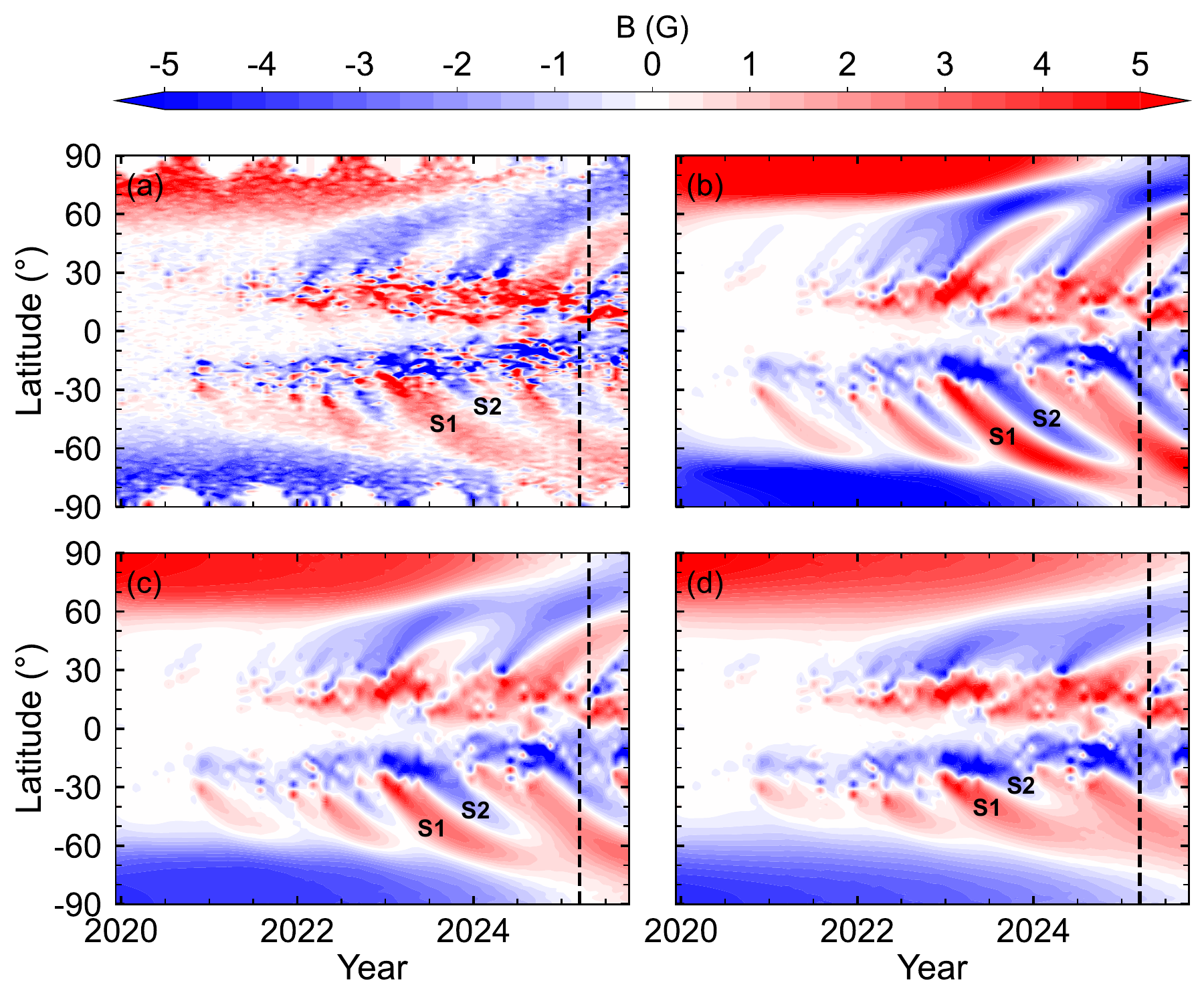}
\caption{Comparison of the surface magnetic field from HMI synoptic magnetograms and the SFT simulations during 2020--2025. 
Panel (a) shows the magnetic butterfly diagram derived from the HMI observations. Panels (b)--(d) show the corresponding SFT simulations using Flow 1--3, respectively. S1 and S2 denote two surges in the southern hemisphere. The dashed lines mark the times at which SO/PHI-HRT observed the respective poles.
\label{fig:butterf}}
\end{figure*}

We first simulate the ongoing Solar Cycle~25 to investigate the influence of the meridional flow on the evolution of the large-scale surface magnetic field. The HMI pole-filled synoptic magnetogram of CR~2225 is adopted as the initial condition, and the surface magnetic field is evolved continuously from CR~2225 to CR~2303, corresponding to a time-frame from December 2019 to November 2025. Figure~\ref{fig:df_pf} compares the axial dipole strength and the high-latitude field, defined as the mean radial magnetic field over the latitude range $60^\circ$--$75^\circ$, derived from the HMI pole-filled synoptic maps and the SFT simulations using the three meridional-flow profiles.

The top panel of Figure~\ref{fig:df_pf} presents the evolution of the axial dipole strength. Since the supergranular diffusivity and flow amplitude of each profile are adjusted to reproduce the same global axial-dipole evolution, the simulated results obtained with the three flows are nearly identical and all agree well with the HMI observations.

The bottom panel of Figure~\ref{fig:df_pf} shows the evolution of the high-latitude field. In contrast to the axial dipole strength, the simulated high-latitude fields exhibit substantial differences among the three meridional-flow profiles. As Flow~2 is optimized against the HMI observations, it successfully reproduces the observed high-latitude-field evolution, including both the timing of the polarity reversal and the amplitude variations. By contrast, the simulation using Flow~1 decreases more rapidly than the observations, leading to an earlier polarity reversal and pronounced fluctuations around and after the reversal. Flow~3 exhibits the opposite behavior: the high-latitude field decreases more slowly than observed, resulting in a delayed polarity reversal and a substantially smoother evolution.

To understand the origin of these differences, Figure~\ref{fig:butterf} compares the magnetic butterfly diagrams derived from the HMI observations and the SFT simulations using the three meridional-flow profiles. Because HMI observes the Sun from a near-ecliptic viewing angle, parts of the polar regions are not visible during many observing periods and therefore appear as blank regions in Figure~\ref{fig:butterf}. Among the three meridional-flow profiles, Flow~2 provides the best overall agreement with the observations. As shown in Figure~\ref{fig:butterf}(c), the simulated polar-field distribution extends from approximately $60^\circ$ to the poles, consistent with the observed pattern. Most of the observed poleward surges are also successfully reproduced. The remaining discrepancies are likely related to limitations in the AR source term and the spatial resolution of the simulations.

In comparison, the simulation using Flow~1, shown in Figure~\ref{fig:butterf}(b), produces poleward surges that are systematically stronger and more concentrated than observed. To illustrate this behavior, we focus on two representative southern-hemisphere surges during 2023--2025, denoted S1 and S2. Surge S1 remains clearly identifiable above $70^\circ$ in the simulation with Flow~1, whereas in the observations it gradually merges with the background polar field above approximately $60^\circ$. The discrepancy is even more pronounced for surge S2. Although only a weak surge is present in the HMI observations, the simulation produces a strong and coherent poleward-migrating feature. As a consequence, the simulated polar field becomes overly concentrated toward the poles and stronger than observed.

Conversely, the simulation using Flow~3, shown in Figure~\ref{fig:butterf}(d), exhibits overly smooth poleward flux transport. The surges become weaker and less distinct than observed: surge S1 appears broader and more diffuse, while surge S2 nearly disappears. Consequently, the simulated polar field is distributed over a wider latitude range and weaker than the observations.

The butterfly diagrams naturally explain the different high-latitude-field evolutions shown in Figure~\ref{fig:df_pf}. The high-latitude field is governed by the poleward transport of magnetic flux from ARs. After June 2022, several prominent surges appear on the solar surface, driving the polarity reversal of the high-latitude fields and the subsequent buildup of the new-polarity fields. Flow~1 produces surges that are stronger than observed, leading to an earlier reversal and larger fluctuations. Conversely, Flow~3 produces weaker surges, resulting in a delayed reversal and a smoother evolution. By accurately reproducing the observed surges, Flow~2 also successfully reproduces the observed evolution of the high-latitude field.

The comparison among simulations using the three flows provides further insight into the role of the meridional-flow profile in shaping the solar polar field. Consistent with earlier studies \citep{DeVore1984, DeVore1987, WangYM2017}, we find that flows reaching maximum speed at higher latitudes produce stronger poleward surges and stronger, more concentrated polar fields. From Flow~3 to Flow~1, the latitude of the flow-speed maximum progressively increases, accompanied by increasingly stronger surges and a progressively greater concentration of magnetic flux toward the poles (Figure~\ref{fig:butterf}). Together with previous studies, these results highlight the critical importance of the meridional-flow profile for both reproducing and interpreting the observed solar polar magnetic field.

It should be noted that the above comparison among different meridional-flow profiles is intended to assess their ability to reproduce the observed polar fields while maintaining a consistent axial dipole evolution. As this comparison required tuning the flow amplitude and supergranular diffusivity for each profile, the resulting differences can not be attributed solely to the peak latitude. To evaluate the importance of these additional parameters, we also test different flow amplitudes and supergranular diffusivities for Flows~1 and 3. These tests show that the parameter adjustments do not substantially improve the agreement with the observed polar fields and instead degrade the reproduction of the axial dipole strength. This result suggests that, despite the influence of other transport parameters, the latitude of the flow-speed maximum is the primary factor responsible for the different simulation results obtained with the three profiles. Future systematic parameter studies will be required to disentangle the respective roles of the peak latitude and other meridional-flow properties in governing the evolution of the solar polar field.

\subsection{Simulation of the SO/PHI-HRT polar-field observations}

\begin{figure*}[htbp]
\centering
\includegraphics[scale=0.7]{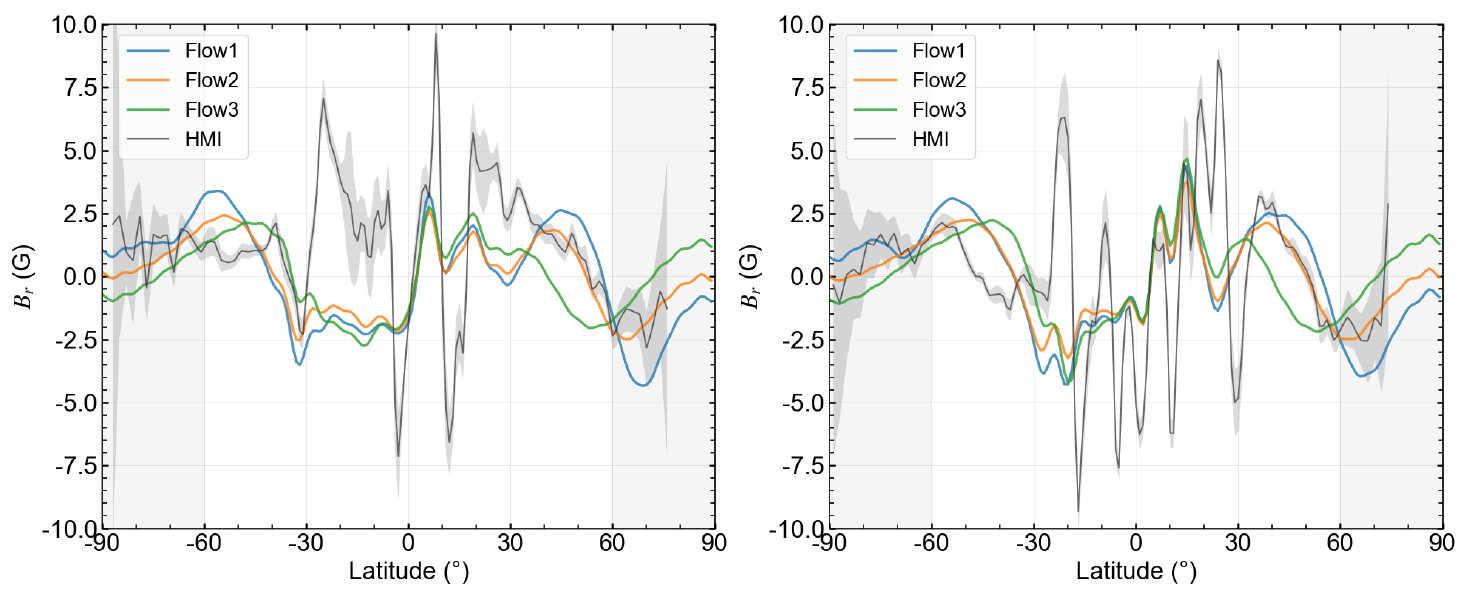}
\caption{Comparison of the latitudinal distributions of the longitudinally averaged radial magnetic field from the SFT simulations and HMI observations. The left and right panels correspond to the northern and southern polar observations by SO/PHI-HRT on 2025 April 24--25 and March 16--17, respectively. The HMI values are obtained by averaging the longitudinally averaged radial magnetic field from multiple full-disk magnetograms during the corresponding SO/PHI-HRT observing periods. The dark gray shading indicates the temporal standard deviation of these longitudinally averaged magnetic fields, while the light gray shading denotes the latitude range $60^\circ$--$90^\circ$.
\label{fig:sml_Br_lat}}
\end{figure*}

\begin{figure}[htbp]
\centering
\includegraphics[scale=0.31]{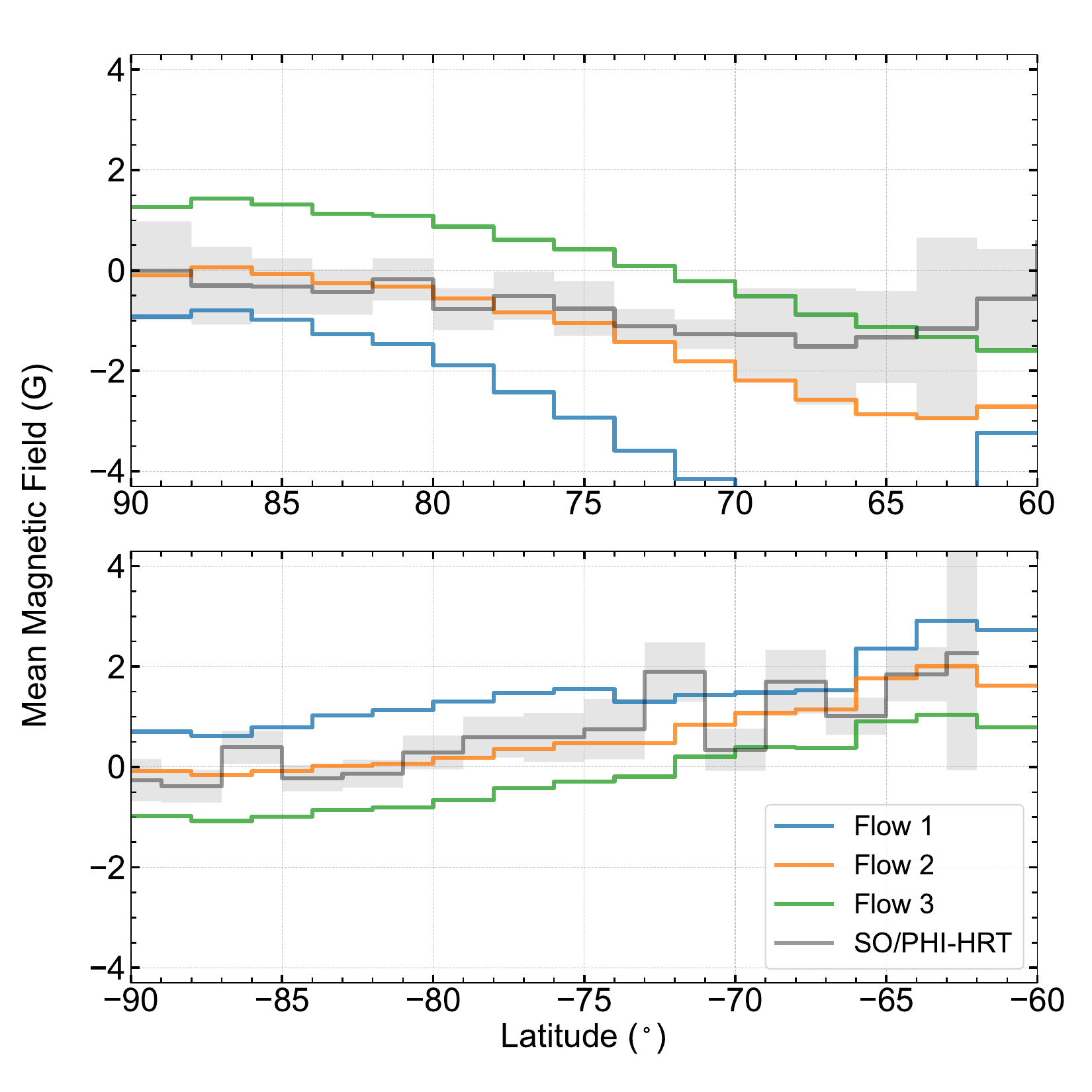}
\caption{Comparison of the longitudinally averaged radial magnetic field between the SFT simulations and the SO/PHI-HRT observations over the latitude range $60^\circ$--$90^\circ$. The top and bottom panels show the northern and southern hemispheres, respectively. The gray shading indicates the temporal standard deviation of the longitudinally averaged radial magnetic field over the corresponding SO/PHI-HRT observing period.}
\label{fig:sml_sophi}
\end{figure}

Until recently, polar-field observations were strongly constrained by the near-ecliptic viewing angle. \citet{Calchetti2025} present the first polar-field measurements obtained by SO/PHI-HRT from outside the ecliptic plane \citep{Gandorfer2018,Solanki2020, Muller2020}. The observation is obtained from heliographic latitudes up to approximately $\pm17^\circ$, about $10^\circ$ higher compared with best near-ecliptic observations, and therefore provide improved spatial coverage and reduced uncertainties in the polar-field measurements. In the following, we use the published SO/PHI-HRT observations to further constrain the SFT simulations.

The SO/PHI-HRT observations of the southern and northern poles were obtained on 2025 March 16–-17 and April 24–-25, respectively. We therefore extract the corresponding two-day intervals from the SFT simulations presented in Section \ref{subs:sml_cycle25} and average the simulated magnetic fields over the same periods. 

Figure~\ref{fig:sml_Br_lat} compares the latitudinal distributions of the longitudinally averaged radial magnetic field from the three simulations with those derived from the SDO/HMI full-disk magnetograms during the corresponding periods. For each latitude, the gray shading indicates the temporal standard deviation of the longitudinally averaged radial magnetic field over the HMI magnetograms obtained during the corresponding observing period. Because the HMI magnetograms only observe the visible hemisphere, whereas the simulations cover the full longitude range, the two are not strictly equivalent, and differences are expected, particularly at low latitudes where the magnetic field is strongly affected by the stochastic emergence of active regions. At higher latitudes, where the longitudinal variation of the field is relatively small, the simulations broadly reproduce the observed high-latitude fields. However, the HMI measurements become increasingly uncertain toward the poles, with large standard deviation above about $75^\circ$ in the south and about $60^\circ$ in the north, and no measurements above about $75^\circ$ in the northern hemisphere. The large standard deviations prevent the HMI observations from clearly distinguishing between the three meridional-flow profiles, as all three simulations fall within the observational standard deviation during these periods. The limited HMI observations at the polar region motivate a comparison with the SO/PHI-HRT observations, which provide improved access to the polar regions. 

% The field above $30^\circ$ is mainly determined by large-scale flux transport. Since the two SO/PHI-HRT observing periods are separated by only about one month, which is short compared with the timescale of poleward flux transport, the typical value of 2.2 year \citep{Wang1991}, the magnetic-field distributions above $30^\circ$ are broadly similar between the two periods. In contrast, the distributions below $30^\circ$ differ substantially because AR emergence is highly variable and stochastic on a timescale of one month.

For the three simulations, the magnetic-field distributions are very similar below $30^\circ$, again reflecting the dominant influence of active-region emergence. However, significant differences appear at higher latitudes. In general, Flow~1 produces a stronger and more concentrated polar field, whereas Flow~3 produces a broader and weaker polar field. The dependence of the polar-field distribution on the meridional-flow profile is consistent with the results reported by \citet{DeVore1987} and \citet{WangYM2017}.

Figure~\ref{fig:sml_sophi} compares the simulated polar fields with the SO/PHI-HRT observations over $60^\circ$--$90^\circ$. The gray shading also indicates the temporal standard deviation of the longitudinally averaged radial magnetic fields during the SO/PHI-HRT observing period. The agreement between the simulations and the observations is quantified by the reduced $\chi^2$, which is 16.94, 1.08, and 5.95 for Flows~1, 2, and 3 in the northern hemisphere, and 6.90, 1.08, and 4.79 in the southern hemisphere, respectively. For a direct comparison, the longitudinal averaging range in the simulations is chosen to match the longitude coverage of the SO/PHI-HRT observations. Owing to the limited spatial resolution of the SFT simulations, the small-scale mixed-polarity flux observed by SO/PHI-HRT cannot be resolved. We therefore only compare the net mean magnetic field.

Compared with the HMI observations shown in Figure~\ref{fig:sml_Br_lat}, the SO/PHI-HRT observations have much smaller standard deviations over $60^\circ$--$90^\circ$, allowing the three simulations to be clearly distinguished. Among them, Flow~2 shows the best agreement with the SO/PHI-HRT observations, with the smallest reduced $\chi^2$ of about 1.08 in both hemispheres. In the southern hemisphere, the simulated magnetic field closely matches the observed values. In the northern hemisphere, the simulation reproduces the observed mean field above $70^\circ$. Although noticeable deviations appear within $60^\circ$--$70^\circ$, the simulated values remain mostly within the observational uncertainties. The remaining discrepancies may arise from uncertainties in the source term and from residual inaccuracies in the transport parameters.

In contrast, Flow 1 and Flow 3 show substantially poorer agreement with the SO/PHI-HRT observations. In the northern hemisphere, Flow 1 produces a strong negative field at $60^\circ$--$90^\circ$, reaching about $-4$ G near $70^\circ$, compared with approximately $-1$ G in the observations. This results in a reduced $\chi^2$ of 16.94. Flow 3 produces a field of the opposite sign and also fails to reproduce the observed near-zero polar field, giving a reduced $\chi^2$ of 5.95. The differences are smaller in the southern hemisphere, but Flow~1 still predicts a positive polar field and Flow~3 a negative one, whereas the observed field remains close to zero near the pole. 
% $-4~\mathrm{Mx\,cm^{-2}}$

The larger discrepancies in the northern hemisphere are likely related to the transient nature of the observed field during the SO/PHI-HRT observing period. The observing times of SO/PHI-HRT are marked in Figure~\ref{fig:butterf}. At this time, a prominent poleward surge is present between $60^\circ$ and $90^\circ$ in the northern hemisphere, whereas the southern hemisphere exhibits only relatively weak surges above $60^\circ$. Surges are relatively difficult features to reproduce in SFT simulations, as discussed in section \ref{subs:sml_cycle25}. Such localized surges may also contribute to the relatively large observational uncertainties of SO/PHI-HRT within the latitude range $60^\circ$--$70^\circ$.

The comparison with the SO/PHI-HRT observations is consistent with the HMI-based results presented in Section~\ref{subs:sml_cycle25}. Flow~1 tends to produce stronger surges and larger polar-field variations, whereas Flow~3 produces weaker surges and smoother polar-field evolution. At the time of the SO/PHI-HRT observations, the observed northern polar field had only recently reversed its polarity. Consequently, the stronger surges produced by Flow~1 lead to an earlier polarity reversal and a stronger negative polar field, while the weaker surges produced by Flow~3 delay the reversal and leave the polar field unrealistically positive. A similar, although weaker, behavior is also seen in the southern hemisphere. 

The SO/PHI-HRT observations provide a strong support of the conclusion that an SFT model with Flow 2 can successfully reproduce the observed large-scale evolution of the solar polar field. The comparison also supports the finding that meridional-flow profiles peaking at higher latitudes produce stronger surges and more rapid variations of the solar polar fields. 
 
% \cite{DeVore1984} show that the steepness of meridional flow with which the velocity declines to zero toward the poles moderate the degree of poleward concentration of the polar field. Greater steepness, the polar field more concentrated.
% \cite{DeVore1987} research the impact of peak latitude and angular breadths. Under the same polarity reversal time, the peak latitude larger, the polar field in the end of cycle is stronger and angular breadths larger (profile is broader), the polar field in the end of cycle is stronger.   no surge, and the simulated polar field vary smoothly.

% \cite{WangYM2017} also compare the impacts of meridional flows peak at middle latitude and near equator, like the Flow 1 and Flow 3 in our paper. Our findings are similar to their results, including meridional flow peak at middle latitude tend to produce stronger surges and stronger polar field with large fluctuations.
% We use three flow further demonstrate the above results. However, \cite{WangYM2017} do not compare the impact of meridional flow in the simulation of real solar cycle. Our results show the flow they used produces too weak surges and also weak polar field with little variations.

\section{Conclusion and Discussion} \label{sec:conclusion}

In this paper, we present a successful SFT simulation of the solar polar magnetic fields observed by SO/PHI-HRT in 2025 March and April during Solar Orbiter's first out-of-ecliptic phases. Using the meridional-flow profile proposed by \citet{WangRH2026}, we successfully reproduce the latitudinal distribution of the longitudinally averaged radial magnetic field observed by SO/PHI-HRT. The same simulation also reproduces the evolution of the axial dipole strength, magnetic butterfly diagram, and high-latitude magnetic fields observed by SDO/HMI during Solar Cycle~25. These results support the proposed meridional-flow profile and demonstrate that an SFT model can reproduce the observed large-scale evolution of the solar polar field, including the polar fields newly accessible to SO/PHI-HRT.

To investigate the origin of this success, we compare simulations using the meridional-flow profile of \citet{WangRH2026} with those employing two representative profiles proposed by \citet{van_Ballegooijen1998} and \citet{WangYM2017}. The three profiles mainly differ in the latitude of their flow-speed maxima. The comparison shows that the meridional-flow profile strongly regulates poleward flux transport and the resulting polar-field evolution. Profiles with higher peak latitudes tend to produce stronger poleward surges and stronger, more concentrated polar fields. These results are consistent with earlier studies \citep{DeVore1984, DeVore1987, WangYM2017}, and further emphasize the importance of accurately determining the meridional-flow profile for realistically reproducing and interpreting the evolution of the solar polar magnetic field.

It should be noted, however, that to maintain a similar evolution of the axial dipole strength while assessing the ability of each profile to reproduce the observed polar fields, both the flow amplitude and the supergranular diffusivity were adjusted for each profile. Consequently, the differences among the simulations cannot be attributed solely to the latitude of the flow-speed maximum. Future systematic parameter studies will therefore be required to quantify the respective roles of the peak latitude and other meridional-flow properties in shaping the solar polar field.

% We also note that \citet{DeVore1987} showed that the angular breadth of the meridional-flow profile can significantly influence the polar-field evolution. The profile proposed by \citet{WangRH2026} is broader than the other two profiles (Figure~\ref{fig:MF}), suggesting that its angular breadth may also contribute to its improved performance. Together, these results further emphasize the importance of accurately determining the meridional-flow profile for realistically reproducing and interpreting the evolution of the solar polar magnetic field.

The success of the meridional-flow profile proposed by \citet{WangRH2026}, however, may depend to some extent on the adopted active-region source. For example, \citet{YangSH2024} obtained reasonable agreement with HMI observation between $55^\circ$--$70^\circ$ using a flow profile similar to that of \citet{van_Ballegooijen1998}, whereas the same class of profiles produces unrealistically strong high-latitude field in our simulations. Although their model includes an equatorward flow above $70^\circ$, its influence on the magnetic field below $70^\circ$ is expected to be limited. The discrepancy between the two studies may therefore be related to differences in the adopted active-region source terms. In particular, the simulated activity belts in \citet{YangSH2024} show some differences from the HMI observations, whereas our simulation reproduce the observed activity belts more closely. This comparison suggests that the evolution of the polar field is jointly determined by both the active-region source term and the transport parameters.

Future observations during solar minimum will be particularly valuable for further constraining the meridional flow. During this phase, the latitudinal distribution of the polar field is determined primarily by the balance between meridional flow and supergranular diffusion, making it much less sensitive to uncertainties in AR emergence \citep{Sheeley1989}. As Solar Orbiter continues to increase its orbital inclination and provide observations from progressively higher heliographic latitudes \citep{Muller2020}, future observations, particularly those obtained during solar minimum, are expected to place substantially stronger constraints on the meridional-flow profile and the transport processes governing the evolution of the solar polar magnetic field.

\begin{acknowledgements}
The research is supported by the National Natural Science Foundation of China (grant Nos. 12425305, 12173005, and 12350004) and China's Space Origins Exploration Program. The SDO/HMI data are courtesy of NASA and the SDO/HMI team. Solar Orbiter is a space mission of international collaboration between ESA and NASA, operated by ESA. We are grateful to the ESA SOC and MOC teams for their support. The German contribution to SO/PHI is funded by the BMWi through DLR and by MPG central funds. The Spanish contribution is funded by AEI/MCIN/10.13039/501100011033/ and European Union ``NextGenerationEU/PRTR'' (RTI2018-096886-C5,  PID2021-125325OB-C5,  PCI2022-135009-2, PCI2022-135029-2) and ERDF ``A way of making Europe''; ``Center of Excellence Severo Ochoa'' awards to IAA-CSIC (SEV-2017-0709, CEX2021-001131-S). The French contribution is funded by CNES. This project has received funding from the ERC (grant agreement No. 101097844 — WINSUN).
\end{acknowledgements}

\bibliography{sample7}{}
\bibliographystyle{aasjournal}

\end{CJK*}
\end{document}